\documentclass{aa}

\usepackage{graphicx}
\usepackage{txfonts}

\usepackage{color}
\definecolor{cite}{rgb}{0.,0.,0.5}
\usepackage[colorlinks=true,citecolor=cite]{hyperref}

\usepackage{natbib}
\usepackage{psfrag}

\def\lsi{LS\,I\,+61\,303}
\newcommand{\ergcms}{\ensuremath{\mathrm{erg\,cm^{-2}\,s^{-1}}}}
\newcommand{\ergs}{\ensuremath{\mathrm{erg\,s^{-1}}}}

\begin{document}
\title{On the origin of correlated X-ray/VHE emission from \lsi}

\titlerunning{On the origin of correlated X-ray/VHE emission from \lsi}

\author{V. Zabalza\inst{1}
      \and
      J.M. Paredes\inst{1}
      \and
      V. Bosch-Ramon\inst{2}}

\authorrunning{Zabalza, Paredes \& Bosch-Ramon}

\institute{Dept.~d'Astronomia i Meteorologia, Institut de Ci\`encies
    del Cosmos (ICC), Universitat de Barcelona (IEEC-UB),  
    Mart\'i i Franqu\`es, 1, E08028, Barcelona, Spain.  
    \email{vzabalza@am.ub.es}
    \and Dublin Institute for Advanced Studies, 31 Fitzwilliam Place, Dublin 2,
    Ireland. 
    }

\date{Received 10 July 2010 / Accepted 16 November 2010}


\abstract
{The MAGIC collaboration recently reported correlated X-ray and very 
high-energy (VHE) gamma-ray emission from the gamma-ray binary \lsi\ during
$\sim$60\% of one orbit. These observations suggest that the emission in these
two bands has its origin in a single particle population.}
{We aim at improving our understanding of the source behaviour by explaining the
simultaneous X-ray and VHE data through a radiation model.}
{We use a model based on a one zone population of relativistic leptonic
particles at the position of the compact object and assume dominant adiabatic
losses. The adiabatic cooling time scale is inferred from the X-ray fluxes.}
{The model can reproduce the spectra and light curves in the X-ray and VHE
bands.  Adiabatic losses could be the key ingredient to explain the X-ray and
partially the VHE light curves.  From the best-fit result we obtain a magnetic
field of $B\simeq0.2$~G, a minimum luminosity budget of
$\sim2\times10^{35}$~\ergs\ and a relatively high acceleration efficiency. In
addition, our results seem to confirm that the GeV emission detected by Fermi
does not come from the same parent particle population as the X-ray and VHE
emission. Moreover, the Fermi spectrum poses a constraint on the hardness of the
particle spectrum at lower energies. In the context of our scenario, more
sensitive observations would allow us to constrain the inclination angle, which
could determine the nature of the compact object.}
{}


\maketitle

\section{Introduction} 

\begin{figure}
    {\normalsize
        \psfragscanon
        \includegraphics{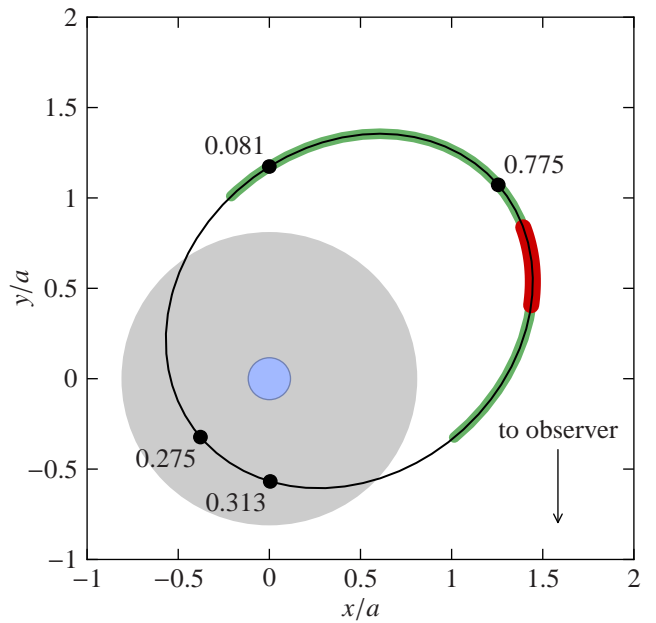}
        \psfragscanoff
    }
    \vspace*{-2em}
    \caption{{Orbital geometry of \lsi\ looking down on the orbital plane
    showing the orbit of the compact object around the Be star in units of the
    orbital semi-major axis. The positions labeled with the orbital phase
    correspond to periastron, apastron, and inferior and superior conjunctions.
    The X-ray/VHE multiwavelength campaign covered the phase range shown in
    green ($0.43<\phi<1.13$), while the phase range of the VHE emission outburst
    is shown in red ($0.6<\phi<0.7$). The Be star (blue circle) and the
    equatorial wind (grey circle) are shown assuming an equatorial disk radius
    of $r_d=7R_\star$. }}
    \label{fig:orbit}
\end{figure}

\object{\lsi}\ is one of the few X-ray binaries (along with
\object{PSR\,B1259-63}, \object{LS\,5039} and
probably Cygnus\,X-1) that have been detected in very high-energy (VHE) gamma
rays\footnote{\object{HESS\,J0632+057} is a variable galactic VHE source thought to be a
binary, but no orbital period has been found yet \citep{hinton09,falcone10}.}.
This source is a high-mass X-ray binary located at a distance of $2.0\pm0.2$~kpc
\citep{frail91} and contains a compact object with a mass between 1 and
4~M$_\odot$ orbiting the main Be star every $26.4960\pm0.0028$~d
\citep{gregory02} in an eccentric orbit \citep[see][and
Fig.~\ref{fig:orbit}]{aragona09}. The main B0\,Ve star has a mass of
$M_\star=(12.5\pm2.5)\, M_\odot$, a radius of $R_\star\approx10\,R_\odot$ and a
bolometric luminosity of $10^{38}$~erg/s \citep{casares05}.  Observations of
persistent jet-like features in the radio domain at $\sim$100~milliarcsecond
(mas) scales prompted a classification of the source as a microquasar
(\citealt{massi04}; see also \citealt{massi09} for a radio spectral discussion),
but later observations at $\sim$ 1--10~mas scales, covering a whole orbital
period, revealed a rotating elongated feature that was interpreted as the
interaction between a pulsar wind and the stellar wind \citep{dhawan06}. In the
X-ray domain, \lsi\ shows an orbital periodicity \citep{paredes97} with
quasi-periodic outbursts in the phase range 0.4--0.8.  The source also shows
short-term flux and spectral variability on time scales of kiloseconds
\citep{sidoli06,rea10}. \lsi\ has been detected in the VHE domain by MAGIC
\citep{magic-lsi} and VERITAS \citep{acciari08}.  It shows a periodic behaviour
\citep{magic-periodic} with maxima occurring around phase 0.6--0.7 and
non-detectable flux around periastron ($\phi_\mathrm{per}=0.275$). However, it
must be noted that since the beginning of 2008 there have been no reports of VHE
detection around apastron even though VERITAS performed several observation
campaigns \citep{holder10}. A recent detection of the source around periastron
\citep{ong10} indicates that its VHE orbital light curve might not be stable
over long periods. Recently
\lsi\ was detected by the high-energy (HE) gamma-ray instrument \emph{Fermi}/LAT
\citep{abdo09}. Its emission at GeV energies was found to be anti-correlated in
phase with the X-ray and VHE emission and presented a spectrum compatible with a
power law and an exponential cutoff at $\sim$$6$~GeV.  This spectrum is
reminiscent of pulsar magnetospheric emission, but no pulsations were found and
no orbital variability would in principle be expected from this kind of
emission. Since March 2009 the source shows a higher phase-averaged flux than
before and the orbital modulation has significantly diminished \citep{dubois10}.
This change in GeV behaviour, as well as the change in the VHE light curve,
indicates that a long term variability effect may be at play. It is not yet
clear, however, whether there is any relation to the $\sim$4.6~yr superorbital
modulation of the radio peak \citep{paredes87,gregory99}.

Two scenarios have been proposed to account for the observational features of
\lsi.  The microquasar scenario requires accretion onto the compact object and
the formation of a jet in which particles are accelerated and emit up to VHE
gamma-rays through upscattering of stellar photons
\citep[e.g.][]{bosch-ramon04,bosch-ramon06}.  In the pulsar wind zone scenario,
the compact object is a young pulsar and particles are accelerated at the
interaction zone between the pulsar wind and the stellar wind
\citep{maraschi81,dubus06psr}. The stellar wind is thought to be formed by a
fast, low-density polar wind and a slow, high-density equatorial decretion disc,
which may be strongly truncated and have its properties disturbed by the compact
object \citep{romero07}. On the other hand, the characteristics of the
interaction of pulsar winds or microquasar jets with a massive star wind are not
well constrained \citep[but see the studies by, e.g,][]{bogovalov08,perucho08}.  
Doppler boosting effects could be present both for a microquasar jet and a
shocked pulsar wind outflow \citep[e.g.][]{khangulyan08-hepro,dubus10}, but its
effects in both scenarios are difficult to assess because of our lack of
knowledge on the geometry and orientation of the emitter.  Finally, the
uncertainty in the orbital geometry, in particular the inclination, also affects
the ability of models to recover the physical parameters of the emitter
\citep{sierpowska09}. All this shows the difficulties of the study of gamma-ray
binaries.

Because of the many factors involved, simultaneous X-ray/VHE observations are a
useful mean to start to disentangle the physics of the high-energy emission in
gamma-ray binaries. However, the combined effect of short-term variability in
the X-ray domain and night-to-night variability in the VHE domain has precluded
a clear detection of X-ray/VHE emission correlation from archival observations.
In 2007, a campaign of simultaneous observations with the MAGIC Cherenkov
telescope and the \emph{XMM-Newton} and \emph{Swift} X-ray satellites revealed a
correlation between the X-ray and VHE bands \citep{magic-correlated}. The
suggestion that the emission in both energy bands comes from the same population
of accelerated particles \citep[see discussion]{magic-correlated} makes these
observations an ideal data set to test the properties of the {X-ray/VHE emitter}
in \lsi.  {It is worth noting that the need of only one leptonic population to
explain the X-ray and the VHE emission would be against lepto-hadronic models
\citep[e.g.][]{chernyakova06}.}

In this work, we present the results of the application of a leptonic one zone
model to explain the measured light curves and spectra from these observations.
Our goal is to obtain the physical parameters of the emitter robustly by making
as few assumptions as possible.

\section{Simultaneous X-ray/VHE observations}
\label{sec:obs}

\lsi\ was observed during  $\sim$60\% of an orbital period simultaneously in
the VHE and X-ray bands in September 2007 \citep{magic-correlated}. During a
first part of the campaign ($0.43<\phi<0.7$), the MAGIC Cherenkov telescope
observed the source for three hours every night and simultaneous
\emph{XMM-Newton} observations were performed.  During a second part
($0.71<\phi<1.13$) the X-ray observations were performed with \emph{Swift}/XRT.
The observation times were of about 15~ks for the \emph{XMM-Newton} observations
and of about 3~ks for the \emph{Swift}/XRT observations. The longer observations
and higher effective area of \emph{XMM-Newton} resulted in much higher
statistics than the measurements from \emph{Swift}/XRT. To take into account
the variability that the source shows on short time scales in the X-ray band when
comparing these fluxes to the VHE measurements, the rms count rate variability
of the X-ray light curve was considered in addition to the statistical
uncertainty in the flux measurement to obtain realistic flux uncertainties
\citep[see][for details]{magic-correlated}.  Using this method, the X-ray flux
uncertainties take values between 6\% and 25\% of the flux. The simultaneous
campaign revealed very similar light curves in the X-ray and VHE bands: a first
outburst at $\phi=0.62$ with a rise time below 20\,h and a decay of about two
days; and a second broader outburst spanning a few days between phases $0.8$ and
$1.1$ with lower peak flux than the first. The correlation coefficient between
the two bands for the first outburst was found to be $r=0.97$, whereas taking
all simultaneous observations lowered the value to $r=0.81$.

Although all X-ray observations resulted in a clear detection, some of the
VHE flux measurements have a significance below 2$\sigma$. For these cases we
will consider the 95\% CL (confidence level) upper limits calculated by
\cite{jogler09}.  The detailed fluxes and uncertainties of the X-ray and VHE
light curves may be found in Tables~1 and 2 of \cite{magic-correlated} and the
VHE upper limits in Table~A.1 of \cite{jogler09}.

The sensitivity of VHE observations precludes obtaining nightly spectra, and
only a spectrum combining the observations with phases between 0.6 and 0.7 was
published by \cite{magic-correlated}. In order to obtain a simultaneous SED, we
extracted an \emph{XMM-Newton} spectrum of the source for the three X-ray
observations that correspond to the MAGIC spectrum. We filtered the data using
SAS~v10.0 and extracted three individual spectra from the pn instrument. We used
the SAS tool \texttt{efluxer} to convert the spectra from counts vs.\ channel
into physical units (flux density vs.\ energy), averaged the three unbinned
spectra and then grouped the bins to a signal-to-noise ratio of 20. In order to
compare the measured spectrum to the computed intrinsic spectrum, we deabsorbed
the former using the \texttt{wabs} FORTRAN subroutine provided with Xspec v12.0
\citep{arnaud96}. We used a column density of $N_\mathrm{H}=5\times10^{21}\
\mathrm{cm}^{-2}$ obtained from the average of the individual fits of an
absorbed power law to the three observations. This column density is
consistent with that of the ISM alone \citep{paredes07}.
 
\section{Model description}
 
The discovery of a correlation between the X-ray and VHE bands points towards a
common mechanism of emission modulation at both
bands. In a leptonic scenario, the fast and simultaneous changes in flux in both
bands indicate that the modulation mechanism has to directly affect the emission
level of the Inverse Compton (IC) and synchrotron processes.  
Assuming constant injection, a way to obtain correlated X-ray and VHE
emission is through a modulation of the number of emitting particles by dominant
adiabatic losses, which would be ultimately related to the
(magneto)hydrodynamical processes in the accelerator and emitter regions. These
processes may be related, for instance, to the interaction of the pulsar wind or
the black hole jet with the stellar wind of the massive companion.  The hard
X-ray spectrum with photon index $\mathrm{\Gamma}\approx1.5$ also points towards
dominant adiabatic losses, which imply an injection electron index of
$\mathrm{\alpha_e}\approx2$.  In the regime of dominant adiabatic losses the
emitted X-ray flux is proportional to the number of emitting particles, so the
orbital dependency of adiabatic losses can be inferred from the X-ray
light curve. Dominant adiabatic losses are not uncommon when modelling gamma-ray
binaries.  They have also been invoked to explain the variations of the X-ray
and VHE fluxes in the gamma-ray binaries PSR~B1259-63 and LS~5039 by
\cite{khangulyan07} (see also \citealt{kerschhaggl11}) and \cite{takahashi09},
respectively. We note that, as explained in Sec.~\ref{sec:icsyn}, the IC
emission will be additionally modulated by the changes in seed photon density
along the orbit as well as the geometrical effects of the location of the
emitter along the orbit.

For the sake of simplicity, and given how little we know about the exact
properties of the particle accelerator and the non-thermal emitter in \lsi, 
we will here assume a constant particle injection spectrum and emitter
magnetic field along the orbit. We will also adopt one leptonic population
cooling down under variable adiabatic losses at the location of the emitter,
which will be approximated as homogeneous and point-like.

The orbital parameters were adopted from \cite{aragona09} with an
inclination angle of $i=45^\circ$. We will also discuss other possibilities
regarding the inclination angle, in particular the extremes of the allowed range
$15^\circ \la i \la 60^\circ$ considered by \cite{casares05}.
{A sketch of the orbital configuration of \lsi\ can be seen in
Fig.~\ref{fig:orbit}  with the phase ranges of the simultaneous X-ray/VHE
campaign indicated. The equatorial wind disk in binaries of the characteristics
of \lsi\
is expected to be truncated at a radius of $r_d=5R_\star$ \citep{grundstrom07}.
However, the variability in the equivalent width of the H$\alpha$ line indicates
that its radius can vary between $4.5R_\star$ and $7R_\star$
\citep{zdziarski10,sierpowska09}. A variable red shoulder in the H$\alpha$ line
\citep{mcswain10} could point towards the creation of a tidal stream in the
equatorial disk because of the proximity of the compact at periastron that
eventually falls back onto the disk at later phases. Even at its largest
possible radius of $7R_\star$, as assumed in Fig.~\ref{fig:orbit}, the
equatorial disk does not affect the phase ranges of the observations we are
considering, and thus we will not include it in our model.}

\subsection{Adiabatic cooling}

One of the remarkable features of the VHE light curve of \lsi\ is the lack of
detectable emission during periastron \citep{magic-periodic}.  IC emission is
very effective at this phase owing to the high stellar photon density, so the
lack of emission means that either it is strongly absorbed or that the number of
accelerated electrons falls drastically. However, the angular dependency of the
pair production process, $\propto (1-\cos\psi)$ for the interaction rate and
$\propto (1-\cos\psi)^{-1}$ for the gamma-ray energy threshold (where $\psi$ is
the interaction angle of the incoming photons), implies that absorption is very
low for phases immediately after periastron even under the dense seed photon
field of the star (see Sec.~\ref{sec:gg} for details on the calculation of pair
production absorption).  Inverse Compton emission also has a dependency on the
angle between the seed photon and the accelerated electron and would be lower
during these phases, but this effect is diminished at higher energies when
scattering takes place in the deep Klein-Nishina regime \citep[see Fig.~6
in][]{khangulyan08}. A reduction in the number of accelerated particles is
required to explain the lack of VHE emission around periastron. Although the
X-ray flux during periastron is around half of the peak flux during the periodic
outbursts, the non-detection of VHE emission during periastron places an upper
limit at only a 10\% of the outburst peak VHE flux. Taking all this into
account, we will assume that in the context of one electron population the
X-ray/VHE correlation is only valid for the X-ray excess flux above a certain
quiescent or pedestal emission. Therefore, the number of emitting particles
responsible for this excess X-ray flux is lower and their VHE emission remains
at non-detectable levels around periastron. This reasoning is supported by the
X-ray/VHE correlation found by \cite{magic-correlated}, in which the constant
term of the linear correlation is $(12.2^{+0.9}_{-1.0})\times10^{-12}\,\ergcms$.
Considering that some of the X-ray observations are below this constant term, we
took the pedestal flux value to be $F_\mathrm{ped}=11.5\times10^{-12}\,\ergcms$.

\begin{figure}
    {\normalsize
        \psfragscanon
        \includegraphics{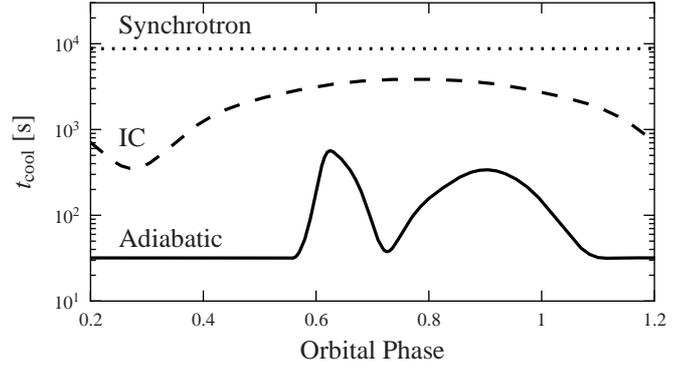}
        \psfragscanoff
    }
    \caption{Cooling times along the orbit for adiabatic (solid), synchrotron
    (dotted) and IC (dashed). The synchrotron and IC cooling times
    are shown at $E_\mathrm{e}=10^{12}$~eV. This energy corresponds to electrons
    that emit in the X-ray band through synchrotron and in the VHE band through
    IC.}
    \label{fig:tcool}
\end{figure}

As mentioned above, a scenario with dominant adiabatic losses is a very good
candidate mechanism to explain the correlation characteristics of the source.
When considering a system with constant magnetic field and constant injected
electron spectrum, the only modulation in synchrotron radiation is due to the
modulation of the dominant cooling process that results in a modulation
of the (evolved) stationary electron distribution.
For dominant adiabatic losses and a constant injection rate, the
X-ray flux will be proportional to the adiabatic cooling time $t_\mathrm{ad}$. A
consistent calculation of $t_\mathrm{ad}$ would require knowledge of the exact
nature of the source and the (magneto)hydrodynamical processes that take place
there. One can however infer $t_\mathrm{ad}$ along the orbit by
taking it proportional to the excess flux above the pedestal
$F_\mathrm{X}-F_\mathrm{ped}$, which is independent 
of the electron energy. For the phases covered by the X-ray observations
we derived a smooth curve following the behaviour of the X-ray fluxes. For
the phases outside the campaign coverage we assumed a flat light curve
consistent with the first four observations at phases 0.43 to 0.59 and the last
observation at phase 0.13.  Choosing the ratio $t_{\rm
ad}/(F_\mathrm{X}-F_\mathrm{ped})$ as small as possible while keeping the
adiabatic losses dominant allows us to put a lower limit on the injected
luminosity budget. 
This choice of the ratio $t_{\rm ad}/(F_\mathrm{X}-F_\mathrm{ped})$
yields adiabatic loss time scales of between 35 and 500 seconds.
Figure~\ref{fig:tcool} shows the orbital evolution of the
cooling time scales inferred for adiabatic and calculated for IC and
synchrotron losses. The adiabatic cooling time is independent of the electron
energy, but for synchrotron and IC the cooling time scales are given at
$E_\mathrm{e}=10^{12}$~eV. This energy corresponds to electrons that emit in the
X-ray band through synchrotron and in the VHE band through IC.

\subsection{Particle energy distribution}
\label{sec:dist}

The radiation emitted by a population of non-thermal particles arises from the
contribution of electrons of different ages. As electrons loose
energy through radiative or adiabatic cooling processes, an evolved particle
energy distribution, different from the injected particle spectrum ($Q(\gamma)$,
taken here constant in time), arises. The evolved particle spectrum
$n(\gamma_\mathrm{e},t)$ can be obtained from $Q(\gamma)$ through the following
equation \citep{ginzburg64}:
\begin{equation}   
      \frac{\partial n(\gamma_\mathrm{e},t)}{\partial t}+\frac{\partial\dot\gamma
      n(\gamma_\mathrm{e},t)}{\partial\gamma_\mathrm{e}}
      =Q(\gamma_\mathrm{e}),
      \label{eq:dist}
\end{equation}
where $\dot\gamma$ is the sum of the energy loss rates for IC, synchrotron and
adiabatic processes and $Q(\gamma_\mathrm{e})$ is the acceleration rate. Since
little is known about the acceleration process responsible for particle
acceleration in \lsi, we assume a constant particle injection spectrum along the
orbit. We adopt a phenomenological electron injection spectrum 
$Q(\gamma_\mathrm{e})$ consisting of a power law function with an exponential
high energy cutoff:
$Q(\gamma_\mathrm{e})=Q_0\ \gamma_\mathrm{e}^{-\alpha_\mathrm{e}}
\exp(-\gamma_\mathrm{e} mc^2/E_\mathrm{e,max} )$.
The maximum electron energy $E_\mathrm{e,max}$ is obtained from the balance of
acceleration and energy loss rates, thus establishing a cutoff in the
accelerated particle energy spectrum (see Sec.~\ref{cutoff});
$\alpha_\mathrm{e}$ is typically $\sim2$ in the context of non-relativistic first order Fermi
acceleration \citep[see, e.g.,][]{drury83}; and $Q_0$ is the normalization of
the function, which we keep constant along the orbit.

For \lsi\ the adiabatic and radiative cooling time scales of
high-energy electrons are much shorter than the orbital period, and thus we can
calculate the steady-state electron energy distribution along the orbit omitting
the first term in Eq.~\ref{eq:dist} and solving for $n(\gamma_\mathrm{e})$:
\begin{equation} 
    n(\gamma_\mathrm{e},\phi) = \frac{1}{|\dot\gamma(\phi)|}
    \int_{\gamma_\mathrm{e}}^{\gamma^\mathrm{max}_\mathrm{e}}
    Q(\gamma')\,\mathrm{d}\gamma', \label{eq:stationary}
\end{equation}
The resulting $n(\gamma_\mathrm{e},\phi)$ is the evolved particle spectrum for
which the emitted radiation properties are calculated along the orbit.

\subsection{Maximum electron energy}

\begin{figure}
    {\normalsize
        \psfragscanon
        \includegraphics{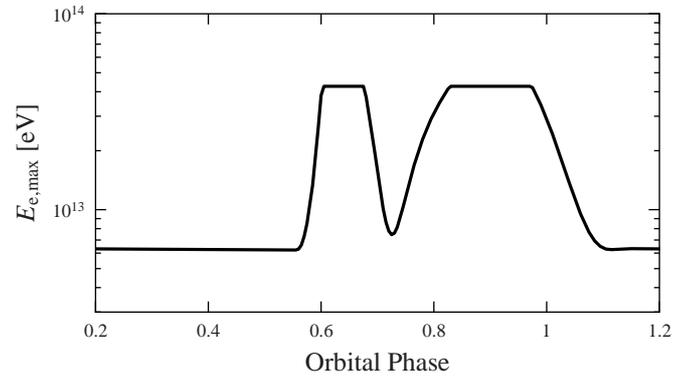}
        \psfragscanoff
    }
    \caption{Maximum electron energy $E_\mathrm{e,max}$ along the orbit taking
    $\eta=10$ (see text for details).}
    \label{fig:emax}
\end{figure}

The VHE gamma-ray energy spectrum of \lsi\ extends up to E$_\gamma\sim10$~TeV
\citep[see][]{acciari09} without a break in the spectrum, indicating that
electrons must be accelerated up to energies beyond 10~TeV. The acceleration of
electrons has to be faster than the cooling times of the different cooling
mechanisms so that the electrons can reach these very high energies. The maximum
electron energy will be that for which $t_\mathrm{acc}=t_{\rm cool}$.
\label{sec:accel}
The acceleration time of electrons can be characterized as
\begin{equation}
t_\mathrm{acc} = \eta r_\mathrm{L} /c \approx 0.11 \eta E_\mathrm{TeV}
B_\mathrm{G}^{-1}\ \mbox{s}\,,
\end{equation}
where $r_\mathrm{L}=E/eB$ is the Larmor radius, $B_\mathrm{G}$ is the magnetic
field in Gauss and $E_\mathrm{TeV}$ is the electron energy in TeV. The
dimensionless constant $\eta\geq1$ indicates the efficiency of the acceleration
process, where $\eta\rightarrow1$ corresponds to the maximum possible
acceleration rate allowed by classical electrodynamics and
$\eta=2\pi(v_\mathrm{s}/c)^{-2}$ for parallel non-relativistic diffusive
shocks in the Bohm regime \citep{drury83}. 

If electron energy losses are dominated by adiabatic cooling, the maximum
energy that electrons will reach is given by $t_\mathrm{acc}=t_\mathrm{ad}$,
which results in
\begin{equation}
    E_\mathrm{e,max}\approx9
    B_\mathrm{G}
    t_\mathrm{ad}
    \eta^{-1}\ \mbox{TeV}\,. \label{eq:ademax}
\end{equation}
However, at some points along the orbit in which adiabatic losses are smaller,
synchrotron losses may dominate at high energies, since $t_{\rm
syn}\propto E_\mathrm{e}^{-1}$ whereas the adiabatic cooling time is constant
with electron energy. In these cases, with a synchrotron cooling time described
by
\begin{equation}
      t_\mathrm{syn} \approx 400\,E_\mathrm{TeV}^{-1}B_\mathrm{G}^{-2}\
      \mbox{s},
\end{equation}
the balance $t_\mathrm{acc}=t_\mathrm{syn}$ results in a maximum energy of 
\begin{equation} 
      E_\mathrm{e,max}\approx 60\,B_\mathrm{G}^{-1/2} \eta^{-1/2}\ \mbox{TeV}\,.
\label{eq:synemax}
\end{equation} 
Inverse Compton energy losses at the relevant energies are smaller than to the
synchrotron and adiabatic losses.  At each point along the orbit, all
relevant time scales are evaluated and the corresponding $E_\mathrm{e,max}$ is
obtained, which in this work characterizes the position of the cutoff in the
injected particle spectrum through an exponential cutoff.  The adiabatic cooling
times inferred from the X-ray light curve range from a few tens to a few hundred
seconds. When considering the requirement that acceleration takes place up to
electron energies of 10~TeV, we can see that the accelerator required must be
quite efficient with $\eta\approx7\mbox{--}130$ depending on the orbital phase.
Since we are assuming that the injection is constant, we also expect the
acceleration rate to be constant and have chosen a value of $\eta=10$ to perform
the energy cutoff calculations along the orbit, {which would correspond to
acceleration in a parallel diffusive shock with velocity of $\sim$$0.8c$ 
(therefore favoring a mildly relativistic shock)}.  The
orbital dependency of the maximum electron energy for $\eta=10$ is shown in
Fig.~\ref{fig:emax}. During periods of low X-ray flux $E_\mathrm{e,max}$ has a
behaviour proportional to $t_\mathrm{ad}$ (see Fig.~\ref{fig:tcool}). During the
outburst peaks, on the other hand, high energy losses are dominated by
synchrotron and the cutoff is constant (following Eq.~\ref{eq:synemax} for
constant magnetic field).
\label{cutoff}

\subsection{Synchrotron and IC emission}\label{sec:icsyn}

For each position along the orbit, we calculate the emitted synchrotron spectrum
from the evolved particle distribution and the magnetic field. Synchrotron
emission in the X-ray range will be generally characterized by a photon index of
$\Gamma_\mathrm{X}=(\alpha_\mathrm{e}+1)/2\approx1.5$. The orbital dependence of
the electron maximum energy will affect the shape of the spectrum, typically
hardening (softening) it for high (low) fluxes, which correspond to lower
(higher) adiabatic losses and therefore a higher (lower) maximum electron
energy. However, since the cutoff in the particle distribution is located at
energies significantly higher than those responsible for X-ray emission, this
effect is small in the X-ray energy band ($\Delta\Gamma_\mathrm{X}\sim 0.2$).

We calculate the IC component of the spectrum using the anisotropic cross
section and the stellar radiation field (blackbody radiation at $kT=2$ eV and
$L_\star=10^{38}$\,\ergs) as the source of seed photons for the interaction.
Following Eqs.~20 and 21 of \cite{aharonian81}, the photon spectrum emitted by a
population of isotropically distributed electrons can be described with a
precision better than 3\% (for $\epsilon\gg\epsilon_0$ and $\gamma\gg1$) by
\begin{equation}
      \frac{\partial^2
      N(\theta,\epsilon)}{\partial\epsilon\partial\Omega}\simeq
      \frac{3}{16\pi}c\sigma_\mathrm{T} \int_{\epsilon_{0,m}(\gamma,\theta)}
      \frac{n_{\epsilon_0}}{\epsilon_0\gamma^2} f(\epsilon,\epsilon_0,\theta,\gamma)
      \mathrm{\,d}\epsilon_0,
\end{equation}
where $\epsilon$ and $\epsilon_0$ are the energies of the scattered and incident
photons, respectively, in units of $m_ec^2$, $\theta$ is the scattering angle,
$\sigma_\mathrm{T}$ is the Thomson cross section, $n_{\epsilon_0}$ is the
number density of the stellar photon field, 
\begin{equation}
      \epsilon_{0,m}(\gamma,\theta)=\frac{\epsilon}{2(1-\cos\theta)\gamma^2[1-\epsilon/\gamma]},
\end{equation}
and
 \begin{equation}
      f(\epsilon,\epsilon_0,\theta,\gamma)=1+\frac{z^2}{2(1-z)}-
      \frac{2z}{b_\theta(1-z)}+\frac{2z^2}{b^2_\theta(1-z)^2},
\end{equation} 
where $b_\theta=2(1-\cos\theta)\epsilon_0\gamma$ and $z=\epsilon/\gamma$. For a
blackbody seed photon distribution, the shape of the resulting emitted spectrum
only depends on the parameter $b_\theta$ and the relativistic electron energy
distribution.
 
\subsection{Pair production absorption}  

The intense radiation field can also absorb very
high energy $\gamma$-ray emission through pair production
$\gamma\gamma\rightarrow\mathrm{e}^+\mathrm{e}^-$ \citep{gould67}. An
exploration of the effects that this absorption can have on the observed
spectrum of gamma-ray binaries can be found in \cite{bosch-ramon09} and references therein. The
differential opacity for an emitted $\gamma$-ray of energy $\epsilon$ is given by
\begin{equation}
      \mathrm{d}\tau_{\gamma\gamma}= (1-\cos\psi) \sigma_{\gamma\gamma} n_{\epsilon_0}
      \mathrm{\,d}l \mathrm{\,d}\epsilon_0,
      \label{eq:difgg}
\end{equation} 
where $l$ is the distance along the line of sight, $\epsilon_0$ is the energy of
the stellar photon, $\psi$ is the interaction angle and the absorption cross
section can be represented in the form:
\begin{equation} 
      \sigma_{\gamma\gamma}=\frac{\pi r^2_e}{2}(1-\beta^2)
      \left[2\beta(\beta^2-2)+(3-\beta^4)
      \ln\left(\frac{1+\beta}{1-\beta}\right)\right],
      \label{eq:ggsig}
\end{equation}
where $\beta=(1-1/s)^{1/2}$, and $s=\epsilon\epsilon_0(1-\cos\psi)/2$. Pair
production can only occur for $s>1$, when the centre of mass energy of the
incoming photons is sufficiently high to create an electron positron pair. The cross
section maximum takes place for $s\simeq3.5$, with a value of
$\sigma_{\gamma\gamma}\simeq0.2\sigma_\mathrm{T}$, and then decreases for
$s\gg1$.

The optical depth owing to pair production $\tau_{\gamma\gamma}$ is calculated at
each point along the orbit by integrating the differential opacity
(Eq.~\ref{eq:difgg}) along the line of sight and over the blackbody seed photon
distribution. The attenuation factor $\kappa=\exp(-\tau_{\gamma\gamma})$ is
applied to the intrinsic IC spectrum (dashed line in Fig.~\ref{fig:sed}) to
obtain the spectrum emitted out of the binary system (solid line in
Fig.~\ref{fig:sed}).
\label{sec:gg}

\section{Results}

As seen in Fig.~\ref{fig:lc}, we were able to reproduce the X-ray and VHE
light curves obtained during the 2007 multiwavelength campaign. Although
the good agreement with the X-ray light curve is expected because we derived
the dominant adiabatic losses from the observed X-ray fluxes, the VHE
light curve is a prediction of the model taking into account the binary geometry,
the stellar photon field, and the non-thermal particle distribution.
Furthermore, the spectral indices in the X-ray and VHE energy bands for the
outburst at phases $0.6<\phi<0.7$ are well reproduced, as can be seen in
Fig.~\ref{fig:sed}.

\subsection{Electron population}

\begin{figure}
    {\normalsize
    \psfragscanon
    \includegraphics{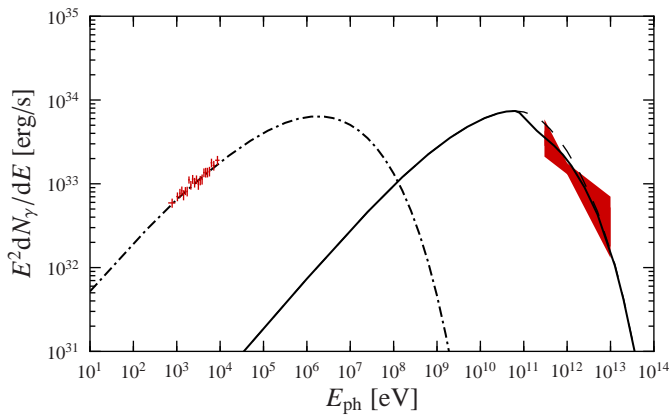}
    \psfragscanoff
    }
    \caption{Computed SED averaged over the three observation periods during the
    first outburst (phases 0.62, 0.66 and 0.70), with synchrotron (dot-dashed), IC
    (solid) and non-absorbed IC (dashed) components. The calculations were
    performed with $i=45^\circ$ and $B=0.22$~G. The crosses show the {\it
    XMM-Newton} EPIC-pn spectrum averaged over three observations and deabsorbed
    taking $\mathrm{N_\mathrm{H}=5\times10^{21}\ cm^{-2}}$. To compare the
    measured flux to the excess computed X-ray flux, we subtracted a
    pedestal power law spectrum corresponding to a $0.3-10$ keV flux of
    $11.5\times10^{-12}\,\ergcms$. The MAGIC simultaneous spectrum is shown as a
    red bow-tie.}
    \label{fig:sed}
\end{figure}

We found that the best fit to the X-ray and VHE photon indices can be
obtained by taking a particle energy distribution with slope
$\alpha_\mathrm{e}=2.1$. As mentioned above, the photon index in the X-ray band
directly depends on the particle index of the parent electron population. In the
X-ray band, the observations show a photon index between $1.58\pm0.02$ and
$1.66\pm0.02$ (plus a $1.85\pm0.02$ outlier) for the \emph{XMM-Newton}
observations, which is anti-correlated with the X-ray flux. The time-dependent
energy cutoff given by the variability of cooling time scales provides a small
time dependent variation of the photon index through a modification of the shape
of the spectrum.  Regarding the VHE observations, the measured photon index is
$2.7\pm0.3_\mathrm{stat}\pm0.2_\mathrm{sys}$ for the three observations
during the first outburst. Because absorption is low at the orbital phases of
the maximum, the photon index is determined by $\alpha_\mathrm{e}$ and the IC
interaction angle.   The parent particle population with $\alpha_\mathrm{e}=2.1$
results in computed X-ray photon indices in the range $1.55$--$1.67$ and
reproduces the observed anti-correlation between $\Gamma_{\rm X}$ and $F_{\rm
X}$.  In the VHE band the adopted $\alpha_{\rm e}$ results in photon indices in
the range $2.6$--$2.8$ for high-flux phases, whereas the spectrum is much softer
for low flux phases because of the lower energy cutoff.  Figure~\ref{fig:sed}
presents the SED averaged over the phase ranges of the three simultaneous
observations during the first outburst (phases 0.62, 0.66 and 0.70), and shows
that both the flux levels and the photon indices at X-ray and VHE are well
reproduced.  

\begin{figure}
    {\normalsize
    \psfragscanon
    \includegraphics{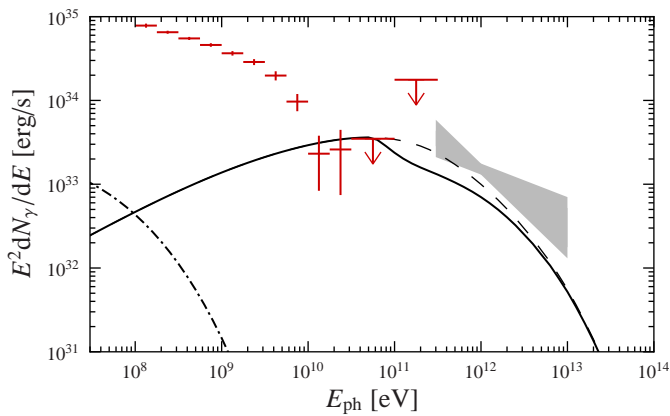}
    \psfragscanoff
    }
    \caption{Computed SED averaged over the whole orbit with synchrotron
    (dot-dashed), IC (solid) and non-absorbed IC (dashed) components, calculated
    with the same parameters as in Fig.~\ref{fig:sed}. The
    crosses and upper limits indicate the phase-averaged \emph{Fermi}/LAT
    spectrum. The MAGIC VHE spectrum during the outburst is shown as a bow-tie.
    It must be noted that the MAGIC spectrum corresponds to different phases
    than the computed SED and is shown as a reference to Fig.~\ref{fig:sed}.}
    \label{fig:sedavg}
\end{figure}

\begin{figure*}
    \sidecaption
    {\normalsize
    \psfragscanon
    \includegraphics{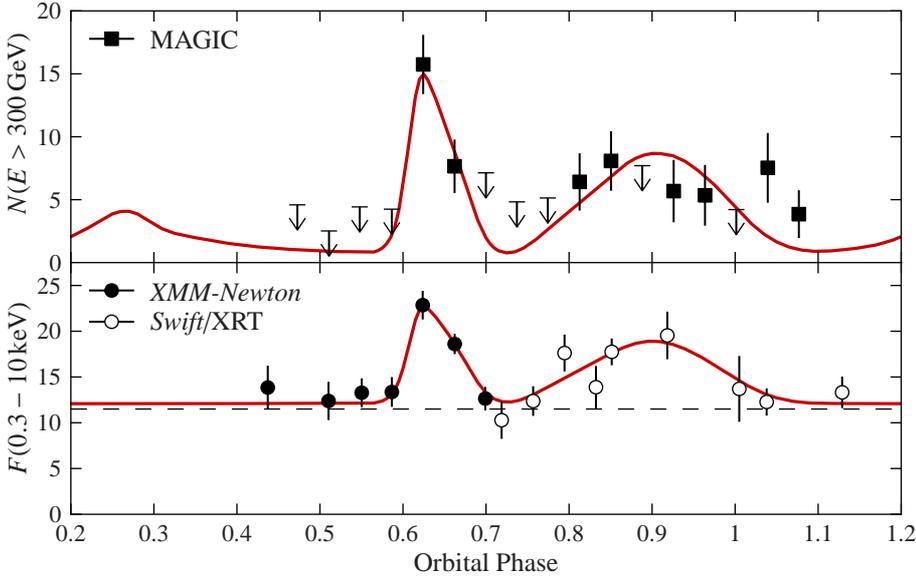}
    \psfragscanoff
    }
    \caption{\emph{Top:} Computed VHE light curve (red line) and observed VHE
    light curve by MAGIC in units of $10^{-12}\,\mathrm{ph/cm^2/s}$.
    Observations with a significance above $2\sigma$ are
    shown in filled squares, while 95\%~CL upper limits are shown otherwise.
    \emph{Bottom:} Computed X-ray light curve (red line) and X-ray light curve
    observed during the multiwavelength campaign with \emph{XMM-Newton} (filled
    circles) and \emph{Swift}/XRT (open circles) in units of
    $10^{-12}\,\mathrm{erg/cm^2/s}$. The pedestal flux is indicated
    by a horizontal dashed line. All error bars correspond to $1\sigma$
    uncertainties. The same parameters as in Fig.~\ref{fig:sed} are used for
    both panels.}\label{fig:lc}
\end{figure*}

We found that the phase-averaged SED calculated using a power law with
$\alpha_\mathrm{e}=2.1$ and a high-energy cutoff produces a flux too high in
energies between 10 and 100\,GeV when compared to the phase-averaged spectrum
measured by \emph{Fermi}/LAT \citep{abdo09}. The GeV spectrum and fluxes along
the orbit appear incompatible with one leptonic population. A way to avoid this
excess flux is to consider the electron injection spectrum described by a broken
power law with a harder particle index below $E_\mathrm{e}=4\times10^{11}$\,eV.
We found that the softest particle index for which the computed GeV
spectrum is compatible with the \emph{Fermi}/LAT measurement is 
$\alpha_\mathrm{e}\la 1.8$. It must be noted that this change in the
particle index at low energies does not affect the spectrum of particles that
emit in the X-ray and VHE bands through synchrotron and IC, respectively. The
\emph{Fermi}/LAT spectrum along with the computed phase averaged SED is shown in
Fig.~\ref{fig:sedavg}.  {The phase-averaged \emph{Fermi}/LAT cannot be easily
compared with the peak spectrum shown in Fig.~\ref{fig:sed} because the
X-ray/VHE and GeV measurements are not simultaneous. Furthermore, the orbital
variability of the GeV spectrum at higher energies (i.e., the shape of the
cutoff) has not been clearly established by \cite{abdo09}, and a phase-resolved
spectrum cannot be derived from a phase averaged-spectrum.}

\subsection{Magnetic field}\label{sec:emprop}

The X-ray/VHE flux ratio is a very sensitive indicator of the magnetic field in
the emitter.  We found the magnetic field that best describes the observed light
curves by performing a exploration of the parameter space of $B$ and the
normalization of the injected particle spectrum $Q_0$. For each $(B,Q_0)$ pair
we calculated the X-ray and VHE light curves and estimated the goodness-of-fit
to the observed flux points using a  $\chi^2$ test. The observed light curves
are best described using an ambient magnetic field of $B=0.22$~G.  It is not
trivial to set a formal uncertainty range on this value because of the unknown
number of constraints that should be considered (e.g., the shape used for the
adiabatic cooling curve). However, we see that for variations of the order of
$0.05$~G around the best-fit value of $B=0.22$~G the observed X-ray and VHE
light curves are no longer well described by the computed ones.

\subsection{Injection energy budget}\label{sec:inj}

The adopted luminosity in injected accelerated particles is around
$2\times10^{35}$\,\ergs\ along the whole orbit. The computed emission bolometric
luminosity, on the other hand, varies between $\sim$$10^{34}$ and
$\sim$$1.5\times10^{35}$\,\ergs. Note that the adopted injected luminosity is only
a lower limit because we arbitrarily chose the lowest ratio between the
X-ray flux and the adiabatic cooling time scale. For higher adiabatic losses
(while still proportional to X-ray emission), the emitted spectrum would be
equivalent but the required injected luminosity would be higher.

The dominant energy band in the broad-band SED of \lsi\ is the MeV-GeV region.
Using the gamma-ray spectrum measured by \emph{Fermi}/LAT \citep{abdo09} we can
obtain a lower limit on the power available for particle acceleration.  The
observed energy flux above 100~MeV is $G_{100}\simeq4.1\times10^{-10}$\,\ergcms,
resulting in an observed luminosity of $2\times10^{35}$\,\ergs. Given this high
luminosity, it is natural to think than the acceleration mechanism at play is
able to provide a similar power to the X-ray and VHE emitting electrons.

\section{Discussion}

\subsection{Adiabatic time scales}

The orbital variability and range of adiabatic losses may be used to pose
constraints on the hydrodynamical properties of the various scenarios proposed
for \lsi. Hydrodynamical calculations have been performed in the binary pulsar
scenario \citep{bogovalov08,khangulyan08-hepro} and for the interaction of a
microquasar jet with the wind of a young star \citep{perucho08,perucho10}.
However, for \lsi\ it is difficult to calculate the adiabatic losses
in each of the mentioned scenarios because of the uncertainties in the properties
of the Be stellar wind and the relativistic outflows. 

In our phenomenological framework, we find that
adiabatic loss time scales range between 35 and 500\,s, as seen in
Fig.~\ref{fig:tcool}. Whereas it is difficult to relate these time scales to
those arising from complex geometries (e.g.~shocked pulsar wind or a microquasar
jet), we can use the simple model of an expanding sphere to obtain insight on
relevant emitter properties. The adiabatic loss time scale $t_\mathrm{ad}$ is
related to the typical scale of the emitter $R_\mathrm{em}$ and its expansion
velocity $v_\mathrm{exp}$ through $t_\mathrm{ad}=R_\mathrm{em}/v_\mathrm{exp}$.
While the size of the emitting region is unknown, the fact that a one-zone
emitter located at the compact object position can successfully explain the
X-ray/VHE emission indicates that the region must be small compared with the
orbital separation.  Considering an emitting region with a radius 10\% of the
orbital separation we obtain a range of radii between $\sim3\times10^{11}$\,cm
(at periastron) and $\sim9\times10^{11}$\,cm (at apastron). For these sizes, the
maximum expansion velocities would be required for the shortest adiabatic
cooling time scale, i.e.~35\,s. At apastron the expansion would need to be highly
relativistic, with a velocity of $0.85c$ (similar to the relativistic sound
speed $c/\sqrt{3}$), corresponding to a Lorentz factor of $\Gamma=1.89$, whereas
for periastron it would be mildly relativistic at $0.29c$.  For the longer
adiabatic loss time scales of hundreds of seconds, the required expansion
velocities would be much lower, of the order of $0.01c$--$0.06c$. 
A change in the size of the emitter could be caused by the change in
external pressure from the stellar wind, resulting in a larger emitter at
apastron \citep{takahashi09}. This would explain the location of the X-ray/VHE
outbursts around apastron, but it is not clear which magnetohydrodynamical
process is responsible for the variation of the adiabatic loss time scale of
more than 
an order of magnitude in 20~h. Wind clumping can be considered as a source of
pressure variability, as suggested by \cite{zdziarski10}, but it would not
explain the constant location of the X-ray/VHE outburst ($0.6<\phi<0.7$) orbit
after orbit. On the other hand, a transition of the emitter from a region
dominated by polar wind to a region dominated by equatorial wind, or vice versa,
could give rise to such a fast change in wind pressure. In addition, there could
be other yet unknown properties of the stellar wind, as shown by the recent
detection of a variable red shoulder in the H$\alpha$ line by \cite{mcswain10},
which could be related to a tidal stream in the equatorial wind falling back onto
the disk. 

\subsection{Energetics}

The energetics of the system, regardless of the acceleration mechanism, are able
to provide the required injected power used in our model given the high
flux in the MeV-GeV band (see Sec.~\ref{sec:inj}). 

The pulsar-like spectrum of \lsi\ in the GeV band (power law with exponential
cutoff at a few GeV) prompted \cite{abdo09} to consider a magnetospheric origin
for the HE gamma-ray emission. In this scenario we can gain some insight on the
spin-down luminosity of the putative pulsar from the HE gamma-ray luminosity.
The observed energy flux $G_{100}$ from magnetospheric emission can be related
to the emitted gamma-ray luminosity $L_\gamma$ as
\begin{equation}
    L_\gamma=4\pi f_{\Omega}(\alpha,\zeta_\mathrm{E})d^2G_{100},
    \label{eq:glum}
 \end{equation}
where $f_\Omega$ is a beaming correction factor that depends on the geometry of
the emitter, the magnetic angle $\alpha$ and the Earth viewing angle
$\zeta_\mathrm{E}$.  The emitted gamma-ray luminosity is related to the pulsar
spin-down luminosity through the gamma-ray efficiency
$\eta_\gamma=L_\gamma/\dot{E}$. A trend was found from EGRET observations that
the efficiency scaled as $\dot{E}^{-1/2}$ \citep{thompson99}, and this was later
confirmed (at least for $\dot{E}>10^{34}$~\ergs) by \emph{Fermi}/LAT arriving to
$\eta_\gamma\simeq 0.034(\dot{E}/10^{36}\ \ergs)^{-1/2}$
\citep{fermi-pulsarcat}. Using this relation, $\dot{E}$ can be inferred from
$L_\gamma$ as $\dot{E}\simeq L_\gamma^2/(1.156\times10^{33}\,\ergs)$. From the
\lsi\ phase-averaged gamma-ray spectrum measured by \cite{abdo09} we obtain
$G_{100}\simeq4.1\times10^{-10}$\,\ergcms, from which
$L_\gamma\simeq2\times10^{35}f_\Omega$\,\ergs. Following these considerations,
the spin-down luminosity inferred from the \emph{Fermi}/LAT observations is of
$\dot{E}\simeq3.3\times10^{37}f_\Omega^2$\,\ergs.  Traditionally (e.g. polar cap
models) it was assumed that the gamma-ray beam covers a solid angle of 1\,sr
uniformly, so $f_\Omega=1/(4\pi)\simeq0.08$, which would result in a spin-down
luminosity of $\dot{E}\simeq2\times10^{35}$\,\ergs. However, more modern pulsar
beaming models (e.g. slot gap and outer gap models) have $f_\Omega\sim1$ for
many viewing angles \citep{watters09}, which would result in a more energetic
pulsar with $\dot{E}\simeq3\times10^{37}$\,\ergs. The fact that the cutoff of
the spectrum is located at high energies ($E_\mathrm{cutoff}=6.3\pm1.1$\,GeV)
points to a high-altitude location for gamma-ray emission, since emission beyond
a few GeV is very difficult close to the neutron star because of strong
attenuation from $\gamma-B\rightarrow e^+e^-$ absorption \citep{baring04}. The
general trend in the \emph{Fermi}/LAT pulsar sample \citep{fermi-pulsarcat} also
points towards a high-altitude location, so we favour slot gap or outer gap
models for which the spin-down luminosity of the putative pulsar in \lsi\ would
be of $\dot{E}\simeq3\times10^{37}$\,\ergs. Even though the \emph{Fermi}/LAT
observations do not completely constrain the spin-down luminosity of the
putative pulsar, the power available for particle acceleration is high. In the
favoured scenario of $\dot{E}\simeq3\times10^{37}$\,\ergs, assuming that the
power in injected electrons constitutes only a fraction $10^{-2}$ of the
spin-down luminosity \citep[e.g.][]{sierpowska08}, the power in relativistic
electron would be larger than our requirement of
$L_\mathrm{inj}=2\times10^{35}$\,\ergs.

Another possibility is that instead of a magnetospheric origin, the emission
detected by \emph{Fermi}/LAT has an Inverse Compton origin. In this case the
efficiency between the spin-down luminosity of the putative pulsar and the
observed IC luminosity is expected to be of 1--10\%. Considering the observed
luminosity of a few times $10^{35}$\,\ergs, the pulsar would need to have a
spin-down luminosity of the order of $10^{37}$\,\ergs.

The higher values of spin-down luminosities pose a problem for the binary pulsar
scenario regarding the shape of the detected radio emission.  \cite{dhawan06}
strongly argued that the rotating cometary-like tail they detected in VLBA
observations was related to the shocked pulsar wind contained by the stellar
wind.  \cite{romero07} already questioned the ability of the stellar wind to
contain the pulsar wind from the results of three-dimensional hydrodynamical
simulations even when considering a significantly lower spin-down luminosity of
$10^{36}$\,\ergs. The shape of the interaction region between the stellar and
pulsar winds is determined by the ratio of wind momentum fluxes
\begin{equation} 
    \eta_\mathrm{w}=\frac{\dot{E}_\mathrm{PSR}}{\dot{M}_\star v_\mathrm{w\star} c},
\end{equation}
where $\dot{M}_\star$ and $v_\mathrm{w\star}$ are the mass-loss rate and wind
velocity of the Be star and $\dot{E}_\mathrm{PSR}$ is the spin-down luminosity
of the pulsar. For the polar wind, we can take
$\dot{M}_\star=10^{-8}\,M_\odot\,\mathrm{yr}^{-1}$ and
$v_\mathrm{w\star}=10^3$\,km\,s$^{-1}$, resulting in a wind momentum flux ratio
of $\eta_\mathrm{w}=0.53(\dot{E}_\mathrm{PSR}/10^{36}\ergs)$ \citep{romero07}.
Ignoring orbital motion (which on the other hand would give rise to small
deviations in the inner region of the interaction, see \citealt{parkin08}), the
opening half-angle of the interaction cone would be
\begin{equation}
    \theta=180^\circ\frac{\eta_\mathrm{w}}{1+\eta_\mathrm{w}},
\end{equation}
as seen in analytical models of wind interaction \citep[e.g.][]{antokhin04}. For
a spin-down luminosity of $10^{36}$\,\ergs\ the stellar wind would barely
dominate the pulsar wind ($\eta_\mathrm{w}=0.53$, $\theta=62^\circ$). 
Considering the favoured high-altitude location for gamma-ray production (i.e.,
slot gap or outer gap models) or an IC origin of the gamma-ray emission, the
pulsar spin-down luminosity is of the order of
$\dot{E}_\mathrm{PSR}\sim10^{37}$\,\ergs. Given this luminosity, the
pulsar wind would dominate the interaction and would not be contained by the
stellar wind. The ratio of wind momentum fluxes would then be of
$\eta_\mathrm{w}\simeq16$ and the half-opening angle would approach $180^\circ$,
posing a completely different scenario from that usually pictured by wind
interaction models \citep[e.g.][]{dubus06psr,sierpowska09,zdziarski10}.  A
lower value of the beaming correction factor $f_\Omega$ in the magnetospheric
scenario could mean that the pulsar spin-down luminosity is as low as
$\dot{E}_\mathrm{PSR}\simeq2\times10^{35}$\,\ergs\ and thus contained by the
stellar wind. However, given the high energy of the cutoff in the \emph{Fermi}/LAT
spectrum of \lsi\ we favour a high-altitude location for gamma-ray production,
leading to the higher value of the spin-down luminosity.

\subsection{Orbital inclination}\label{sec:inc}

Synchrotron emission will have no dependence on the orbital inclination because we
are assuming isotropic particle distribution and a fully disordered magnetic
field. For IC emission, on the other hand, there is a dependence with the angle
$\theta$ between the line of sight and the direction of propagation of the seed
photons (i.e., the connecting line between the star and the compact object).
Given the orbital configuration of \lsi\ {(see Fig.~\ref{fig:orbit})}, 
$\theta$ will not vary significantly with inclination for the phases around
apastron, at which the source is detected in VHE. For $\gamma\gamma$ absorption,
on the other hand, the dependency is on the angle between the directions of the
VHE photon and the seed photon integrating along the line of sight out of the
system.  This means that absorption will show weaker orbital variability for low
inclinations and stronger for high inclinations. For the flux peak observed at
$\phi=0.62$, the optical depth of $\gamma\gamma$ absorption is practically
independent of inclination at $\tau_{\gamma\gamma}\approx0.2$ for 500~GeV
photons. However, the second, wider outburst at phases 0.8--1.1, does have an
absorption dependency with inclination, with $\tau_{\gamma\gamma}\approx1.4$ for
$i=60^\circ$ but $\tau_{\gamma\gamma}\approx0.3$ for $i=15^\circ$. The
diminished absorption at low inclinations leads to a VHE flux higher than the
observed values, particularly taking into account the upper limit measurement at
phase 0.8.  However, given the high uncertainties of the VHE nightly fluxes, the
difference between the light curves computed for $i=60^\circ$ and $i=15^\circ$ is
of less than $1\sigma$. Even though we do not consider it significant given the
present sensibilities, deeper VHE observations with firm detections at these
phases could help to constrain the inclination of the orbit.

\subsection{Magnetic field}

The X-ray/VHE fluxes can be explained using a constant magnetic field of
$B=0.22$~G. A constraint on the accelerator maximum energy can be put by
requiring that the Larmor radius of the accelerated electron
($r_\mathrm{L}=E_\mathrm{e}/qBc$) is contained within the accelerator size. This
translates to a maximum energy of $E_\mathrm{max}=300\,B_\mathrm{G}R_{12}$\,TeV,
where $B_\mathrm{G}$ is the magnetic field in Gauss and $R_{12}$ the accelerator
radius in units of $10^{12}$\,cm. For the magnetic field we found and
considering an accelerator size of the order of 10\% of the orbital separation,
the maximum energy is well above the required 10~TeV and radiative losses
dominate at high energies (see Sec.~\ref{sec:accel}).  In the binary pulsar
modelling approach of \cite{dubus06psr}, the magnetic flux was considered
variable along the orbit as a consequence of the variation in the distance of
the wind-shock region to the pulsar. In the shocked wind region, the magnetic
field values obtained were as high as 10~G. {\cite{chernyakova06} obtain a very
similar magnetic field to ours, with $B=0.25$~G ($B=0.35$~G) for the low (high)
flux state of the source. However, the model is based on an emitting particle
spectrum very soft at high energies, resulting in a synchrotron component
peaking in the radio band and an IC component dominating from soft X-ray to the
MeV band. They require another hadronic component to explain the VHE emission.
The magnetic field in this model traces the flux ratio between radio and X-ray
fluxes instead of ours, which traces the flux ratio between X-ray and VHE.} On
the other hand, in the microquasar approach used by \cite{bosch-ramon06lsi}, the
magnetic field at the base of the jet was taken as 1~G. Most of these approaches
use higher magnetic fields than the one we found to best reproduce the
observational features. An explanation for this is likely the need in these
models to generate both the pedestal and X-ray flux that we considered
independently.

\section{Summary and concluding remarks}

We presented a radiation model of a single leptonic population that can
successfully describe the data obtained through a simultaneous X-ray/VHE
campaign on \lsi\ performed by MAGIC in 2007. The observed X-ray/VHE correlation
indicates both that the emission at these bands may come from a single particle
population and that energy losses may be dominated by adiabatic losses. By
assuming a constant magnetic field and injection along the orbit, we
inferred the adiabatic loss rate from the X-ray light curve and the time
dependent electron maximum energy as the balance between acceleration and energy
loss time scales. We found that a quite efficient accelerator ($\eta\sim10$)
is required to obtain the observed VHE spectra because of the fast adiabatic
cooling. The injected electron spectrum is constant along the orbit and
initially taken as a power law with a high-energy cutoff, but the
\emph{Fermi}/LAT HE spectrum poses a constraint on the hardness of the spectrum
at lower energies, which we estimate has to be harder than
$\alpha_\mathrm{e}\simeq1.8$ for electron energies below $4\times10^{11}$~eV. At
higher energies, on the other hand, an index of $\alpha_\mathrm{e}=2.1$ matches
the observed X-ray and VHE photon indices.  The observed light curves are best
reproduced using a magnetic field of $B=0.22$~G. The general picture shows that
the observed emission in X-ray and VHE is fully compatible with originating
in the same parent particle population under dominant adiabatic losses.

The required luminosity budget in injected electrons for the computed X-ray and VHE
emission is $2\times10^{35}$~erg/s. Both the mi\-cro\-qua\-sar and binary pulsar
scenarios are able to provide these levels of luminosity in accelerated
electrons. The GeV luminosity of \lsi\ can also give us hints on the available
power: independently of whether the detected emission is magnetospheric or IC
emission, the injected power should be $\ga 10^{37}$~erg/s. If it is
magnetospheric emission, the high-altitude favoured scenario of gamma-ray
production implies that the spin-down luminosity of the pulsar is
$\dot{E}=3\times10^{37}$~erg/s.  This high spin-down luminosity poses problems
for the binary pulsar scenario because the pulsar wind would not be contained by the
stellar wind, which would give rise to a completely different scenario from the one
usually considered.

The VHE and X-ray light curves of LS\,5039 and \lsi\ seem relatively similar,
both peaking close to the apastron. For LS\,5039 the maximum is also close to
the inferior conjunction of the compact object, whereas this is not the case in
\lsi.  If the light curve maxima are related to apastron rather than to the
observer orientation with respect to the system, it may point to a similar
physical origin for the X-ray and VHE modulation in both sources.  On the other
hand, the adiabatic losses could have a different cause.  Recall the very
different stellar winds in both sources.

We found that the X-ray and the VHE emission are consistent with a single
parent particle population, whereas the HE gamma-ray emission detected by
\emph{Fermi}/LAT probably has a different origin. The orbital anti-correlation
and the exponential cutoff spectrum suggest that several non-thermal emitting
regions may be present in the system. This could also explain the presence of a
pedestal emission in X-ray that is compatible with constant emission along the
orbit.

We conclude noting that simultaneous observations at the X-ray and VHE bands
provide a unique diagnostic tool to probe the emitter properties in galactic
variable sources.  Future observations, in particular by more sensitive
next-generation VHE observatories, will allow us to better understand these
peculiar systems.

\begin{acknowledgements} 
      We would like to thank Felix Aharonian and Dmitry Khangulyan for useful
      comments.  We also thank the anonymous referee for constructive comments
      and suggestions that helped improve the manuscript.  The authors
      acknowledge support of the Spanish MICINN under grant AYA2007-68034-C03-1
      and FE\-DER funds. V.Z.~was supported by the Spanish MEC through FPU grant
      AP2006-00077. V.B-R.\ thanks the Max Planck Institut f\"ur Kernphysik for
      its kind hospitality and support. V.B-R.\ also acknowledges the support of
      the European Community under a Marie Curie Intra-European fellowship.
\end{acknowledgements}

\bibliographystyle{aa}
\bibliography{15373}

\end{document}